\documentstyle[sprocl,epsfig]{article}

\def\Journal#1#2#3#4{{#1} {\bf #2}, #3 (#4)}

\def\be{\begin{equation}}
\def\ee{\end{equation}}
\def\bea{\begin{eqnarray}}
\def\eea{\end{eqnarray}}
\begin{document}

\title{
DINUCLEAR CONCEPT -- CLUSTER MODEL OF FUSION
}

\author{
G.G.ADAMIAN$^{1,2,3}$, N.V.ANTONENKO$^{1,2}$,
E.A.CHEREPANOV$^{2}$, A.K.NASIROV$^{2,3}$,
W.SCHEID$^1$, V.V.VOLKOV$^{2}$\\
}

\address{
$^{1}$Institut f\"ur Theoretische Physik,
%der Justus--Liebig--Universit\"at,
D--35392 Giessen, Germany \\
$^{2}$Joint Institute for Nuclear Research, 141980 Dubna, Russia\\
$^{3}$Institute of Nuclear Physics,
Tashkent 702132, Uzbekistan
}

\maketitle\abstracts{
The synthesis of superheavy elements is analysed within
the dinuclear system concept of compound nucleus formation.
The perspectives for using radioactive beams in complete
fusion reactions are discussed.}

The existing fusion models are distinguished by the choice
  of the relevant collective degree of freedom which is mainly
  responsible for the complete fusion. For example, many models
  assume a melting of the nuclei along the relative distance.
  It was demonstrated that the adiabatic scenario of fusion in
  the relative distance leads to an overestimation of the fusion
  probability $P_{CN}$ \cite{1} and mostly gives an incorrect isotopic
  trend of $P_{CN}$. In the dinuclear system (DNS) concept \cite{2} the
  compound nucleus is reached by a series of transfers of nucleons
  from the light nucleus to the heavy one. The dynamics of the DNS
  is considered as a combined diffusion in the degrees of freedom
  of the mass asymmetry $\eta=(A_1-A_2)/(A_1+A_2)$ ($A_1$ and $A_2$
  are the mass numbers the DNS nuclei) and of the relative distance
  describing the formation of the compound nucleus and the quasifission
  process (decay of the DNS), respectively \cite{3}. The competition
  between the complete fusion and quasifission processes is
  taken into consideration in the DNS model and leads to a strong
  reduction of the fusion cross section \cite{3,4}. This cluster fusion
  model is justified by the structural forbiddenness effect \cite{5}
  which hinders the nuclei to melt together along the relative
  distance. In the DNS concept  \cite{3} the evaporation residue cross
  section is calculated as $ \sigma_{ER}=\sum_{J}^{}\sigma_c
  P_{CN}W_{sur}$, where $\sigma_c$ is the capture cross section for
  angular momentum $J$. The stabilizing shell effects of the formed
  superheavy compound nucleus against fission in the de-excitation
  process are thoroughly studied by the theory and surviving probabilities
  $W_{sur}$ of compound nuclei are derived. The dependence of the
  probability of complete fusion $P_{CN}$ on nuclear structure effects
  during the fusion process starting from the entrance channel and
  ending with the compound nucleus formation is the crucial factor
  for the correct calculation of $\sigma_{ER}$.
  In the reactions  $^{90}$Zr+$^{90,92,96}$Zr, $^{90,96}$Zr+$^{100}$Mo,
     $^{86}$Kr+$^{99,102,104}$Ru, $^{90,92,94,96}$Zr+$^{124}$Sn and
     $^{86}$Kr+$^{130,136}$Xe the fusion probabilities are decreased \cite{6}
     when the neutron number of projectile or target deviates from the
     magic number. In the DNS model this behaviour is simply explained
     taking the deformation of nuclei and shell effects in dependence
     of the DNS potential energy on $\eta$ into account. For example,
     the value of the energy threshold for fusion, which determines $P_{CN}$,
     is larger in the $^{86}$Kr+$^{130}$Xe reaction than in the
     $^{86}$Kr+$^{136}$Xe reaction.
     So, the values of $P_{CN}$ and $W_{sur}$ are larger
     in the reaction with $^{136}$Xe than with $^{130}$Xe, which leads to a
     difference of the about 3 orders of magnitude in $\sigma_{ER}$ in these
     reactions. In the fusion reactions leading to actinides, for example
     the $^{66,76}$Zn+$^{174}$Yb reaction, the increase of $W_{sur}$ with
     increasing neutron number of the system is stronger than the decrease of
     $P_{CN}$. This gives evident benefit to the neutron-rich projectiles for
     producing actinides.

In contrast to other models, the optimal excitation energy $E_{CN}^*$
     of the compound nucleus formed in cold fusion reactions is reproduced
     in the DNS concept. The value of $E_{CN}^*$ increases after $Z$=113
     (Fig.~1a).
     The difference between the $Q$-values of Refs.~\cite{8,9} for elements
     till $Z$=113 is small. The strong decrease (few orders of magnitude) of
     the cold fusion cross section with increasing charge number $Z$ of the
     compound nucleus \cite{7} is mainly caused by a decrease of the fusion
     probability $P_{CN}$ due to a strong competition between complete fusion
     and quasifission in the DNS (Fig.~1b). Therefore, in reactions
     $^{76,74}$Ge+$^{208}$Pb $\to ^{283,281}$114+1$n$
     we expect a value of $\sigma_{ER}$ which
     is smaller than 0.2 pb. The $\sigma_{ER}$ for the $Z$=116 and 118 elements
     formed in the $^{80,82}$Se,$^{84,86}$Kr+$^{208}$Pb reactions are smaller
     than the value for $Z$=114. In actinide-based reactions
     $^{48}$Ca+$^{232}$Th,
     $^{238}$U, $^{242,244}$Pu, $^{248}$Cm, $^{249}$Cf, the $P_{CN}$ also
     decrease with increasing $Z$, but they are larger than in Pb-based
     reactions.
For $^{48}$Ca+$^{244}$Pu$\to ^{289}$114+3$n$,
     the $P_{CN}$ is 6$\times 10^{-4}$ which is about $10^{5}$ times larger
     than in $^{76}$Ge+$^{208}$Pb$\to^{283}$114+1$n$. The gain in fusion and
     capture probabilities for actinide-based reactions with respect to cold
     fusion reactions is not compensated by loss in the survival probability
     of the compound nucleus. So, our comparison of the formation cross section
     of element $Z$=114 in Pb- and actinide-based reactions shows that the latter
     one is larger ($\sigma_{3n}$=1.5 pb) \cite{4}. The $\sigma_{ER}$ of the
     $^{48}$Ca+$^{248}$Cm,$^{249}$Cf reactions are smaller than experimental
     $\sigma_{ER}$=1 pb of the $^{48}$Ca+$^{244}$Pu reaction.

     In the Pb-based reactions with neutron-rich nuclei $^{70,74,78}$Ni,
     $^{80}$Zn,$^{86}$Ge and $^{92}$Se the increase of $W_{sur}$ with the
     number of neutrons could be compensated by decreasing $P_{CN}$.
For example, in the
     $^{62}$Ni+$^{208}$Pb reaction the yield of the $Z$=110 element is
     comparable with the yields in the $^{70,74}$Ni+$^{208}$Pb reactions.
     The larger lifetime of the neutron-rich superheavies will allow a detailed
     study of their properties.

    One of the authors (G.G.A.) is grateful to the
    Alexander von Humboldt-Stiftung for the support. This work was supported
    in part by DFG and RFBR.

\begin{figure}
\phantom{a}
\vspace*{-0.7cm}
\centerline{\epsfig{figure=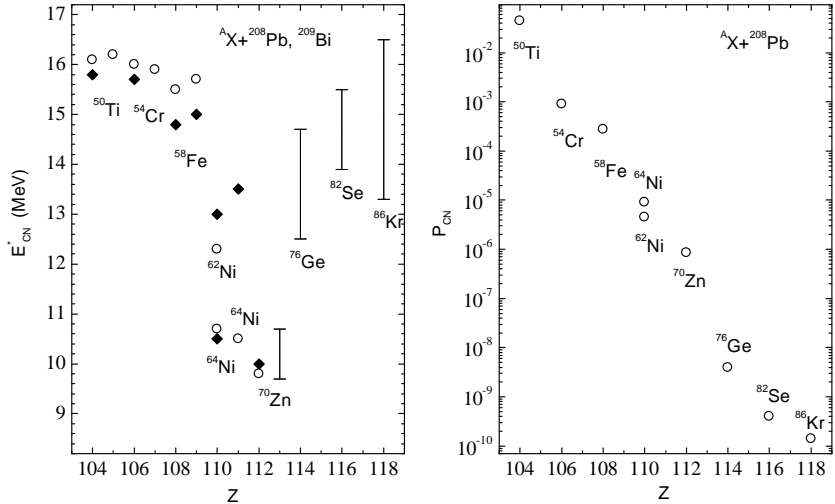,width=12.0cm}}
\vspace*{6.5cm}
\caption{ a) Optimal excitation energies of the compound nuclei.
     b) Calculated fusion
     probabilities $P_{CN}$ for cold fusion (HI,1$n$) reactions.
     The experimental
     data \protect\cite{7} are shown by solid diamonds.
     The projectiles are indicated.
     For compound nuclei with $Z$=104-112, the calculations
     were performed with
     $Q$-values from Ref.\protect\cite{8}.
     For the elements with $Z\ge$113 the lower (upper)
     limit of bars was calculated with $Q$-values from
     Ref.\protect\cite{8}
     (Ref.\protect\cite{9}).}
\vspace*{-0.5cm}
\end{figure}


\section*{References}
\begin{thebibliography}{99}
\bibitem{1} G.G.Adamian et al.,
\Journal{{\em Nucl. Phys.} A}{646}{29}{1999}.
\bibitem{2} V.V.Volkov, \Journal{\em Izv. AN SSSR ser. fiz.}{50}{1879}{1986}.
\bibitem{3} N.V.Antonenko et al., \Journal{{\em Phys. Lett.} B}{319}{425}{1993};
\Journal{{\em Phys. Rev.} C}{51}{2635}{1995};
G.G.Adamian et al.,
\Journal{{\em Nucl. Phys.} A}{618}{176}{1997};
A {627,} {332} {(1997)};
A {633,} {154} {(1998)}; R.V.Jolos et al.,
\Journal{{\em Europ.Phys.J.} A}{4}{245}{1999}.
\bibitem{4} E.A.Cherepanov, \Journal{\em Pramana} {23}{1c}{1999};
Yu.Ts.Oganessian et al., Preprint JINR, E7-99-53 (1998).
\bibitem{5}
G.G.Adamian et al., \Journal{{\em  Phys. Lett.} B}
{451}{289}{1999}.
\bibitem{6} K.H.Schmidt, W.Morawek, \Journal{\em Rep. Prog. Phys.}{54}{949}{1991}.
\bibitem{7} S.Hofmann, \Journal{\em  Rep. Prog. Phys.}{61}{570}{1998};
G.M\"unzenberg \Journal{{\em  Phil. Trans. R. Soc. Lond.} A}{356}{2083}{1998}.
\bibitem{8} P.M\"oller, J.R.Nix, \Journal{\em At. Data Nucl. Data Tables}
{39}{213}{1988}.
\bibitem{9} P.M\"oller et al., \Journal{\em At. Data Nucl. Data Tables}
{59}{185}{1995}.
\end{thebibliography}
\end{document}